\documentclass[aps,prb,twocolumn,superscriptaddress,showpacs,floatfix]{revtex4}

\usepackage{bm}
\usepackage{amsmath}
\usepackage{amssymb}
\usepackage{graphicx}
\usepackage{color}
\usepackage{soul,xcolor}
\setstcolor{red}
\usepackage{ulem}
\usepackage{cancel}

\newcommand\Ccancel[2][red]

\usepackage{xcolor}\definecolor{ao}{rgb}{0.0,0.0,1.0}

\definecolor{br}{rgb}{1.0, 0.22, 0.0}

\usepackage{epsfig}

\newcommand{\up}{\uparrow}
\newcommand{\down}{\downarrow}

\def\sig{{\mbox{\boldmath{$\sigma$}}}}

\input{epsf}
\begin{document}
\title{Spin precession
in spin-orbit coupled weak links:\\
Coulomb repulsion and Pauli quenching}
\author{R. I. Shekhter}
\affiliation{Department of Physics, University of Gothenburg, SE-412
96 G{\" o}teborg, Sweden}

\author{O. Entin-Wohlman}
\email{oraentin@bgu.ac.il}
\affiliation{Raymond and Beverly Sackler School of Physics and Astronomy, Tel Aviv University, Tel Aviv 69978, Israel}
\affiliation{Physics Department, Ben Gurion University, Beer Sheva 84105, Israel}

\author{M. Jonson}
\affiliation{Department of Physics, University of Gothenburg, SE-412
96 G{\" o}teborg, Sweden}

\author{A. Aharony}
\affiliation{Raymond and Beverly Sackler School of Physics and Astronomy, Tel Aviv University, Tel Aviv 69978, Israel}
\affiliation{Physics Department, Ben Gurion University, Beer Sheva 84105, Israel}

\date{\today}

\begin{abstract}
A simple model for the transmission of pairs of electrons through a weak  electric link in the form of a nanowire
made of a material with strong electron spin-orbit interaction  (SOI)  is presented, with emphasis on the effects of Coulomb interactions and the Pauli exclusion principle.
The constraints due to the Pauli principle  are shown to ``quench" the coherent SOI-induced precession of the  spins when the spatial wave packets of the two electrons overlap significantly. The quenching, which results from  the
projection of the
pair's spin states
onto spin-up
and spin-down states on the link, breaks up the coherent propagation
in the link into a sequence of coherent hops that add incoherently. Applying the model  to the transmission of Cooper pairs between two superconductors,
we find that in spite of Pauli quenching,  the Josephson current
oscillates with the strength of the SOI,  but  may even change its sign (compared to the limit of Coulomb blockade, when the quenching is absent). Conditions for an experimental detection of these features
are discussed.

 \end{abstract}


\maketitle

\noindent{\bf Introduction.} Electronic devices comprising  weak electric links made of a material with strong electronic  spin-orbit interaction  (SOI)  are
promising candidates for spintronics applications and  offer new modes of operation. This follows from the possibility
to manipulate the spin structure of electrons
flowing through the link
via the effect of external fields on the
SOI. Transport experiments  on electrons in  gated semiconductor heterostructures  demonstrated
that the strength of the Rashba \cite{Rashba.1960,Bychkov.1984}  SOI  can be both significant and
controlled by
gate voltages. \cite{Nitta.1997,Sato.2001,Beukman.2017} Theoretically,  it was proposed that nanowire weak
links made of SOI-active materials
suspended between bulk electrodes can act as ``Rashba
spin-splitters" and  lead to mechanically controlled spin-selective transport \cite{Shekhter.2013} and spintro-voltaic
effects. \cite{Shekhter.2014}
These
predictions
can be understood within
a semiclassical picture: as the electrons pass through the
link, their spins precess
around an effective magnetic field associated
with the
SOI. This spin dynamics splits
the electron wave function into different
spin states and leads to a certain probability, which can be controlled externally, for the spins to be flipped as they
emerge from the
link. \cite{Shekhter.2016x}

The
spin-splitting phenomenon becomes more complicated for
the transmission of {\it a pair of electrons} through an SOI-active  link.
Weak superconductivity, established  by the transfer of Cooper pairs through a non-superconducting material,  is the obvious  system for studying spin-splitting in the transmission of electron pairs; it is analyzed in this paper.
Other examples of  pairs tunneling appear in non-superconducting transport
involving higher-order tunneling events:  shot noise in weak electric links \cite{Blanter2000}
and cross-correlations of currents
in
multi-terminal mesoscopic structures. \cite{Marcus2007}

Since two electrons in the same spin state cannot occupy the same place simultaneously (Pauli principle),
 possible effects due to   the Coulomb blockade
of electron tunneling
and the constraints on the spin dynamics have to be considered. A fundamental question
is whether these constraints destroy the coherent spin precession of individual electrons and destroy
the spin-splitting phenomenon. Here we estimate the outcome of these effects.  As a motivation for the structure of our model,
we propose a
semiclassical picture of
the  transfer of a pair of electrons through the SOI-active weak link. The two electrons tunnel
one by one at separate times, say $t_1 < t_2$,  from the source electrode to the nanowire.
Once there,  they propagate as wave packets with different velocities until leaving  the nanowire by tunneling into the drain electrode, also at different times,  $t_3 < t_4$.
If the first electron to enter leaves before the second  comes in,  i.e.,   $t_1 < t_3<t_2<t_4$, then there is
only one electron in the wire at any given time. This case
 is  termed
``single-electron", or  ``s-channel" -- transmission    (though, importantly, the transmission is that  of a pair of electrons that traverse the link one by one).
Effects due to the Coulomb interaction
and the
Pauli principle  on the motion
are ignored; the spatial motion is
that of free electrons and the spin dynamics evolve coherently according to the SOI on the link.

If, on the other hand, the second electron tunnels into the nanowire before the
first
one has left, i.e.,   $t_1 < t_2<t_3<t_4$, then there are  two electrons in the
nanowire during the time interval $\{t_{2},t_{3}\}$.
This case is called  ``double-electron", or ``d-channel" -- transmission.
Because of the different longitudinal velocities of their wave packets, the two electrons may or may not meet  somewhere on the wire. If they do, then the constraints due to the Pauli principle
need to be considered.
This is accomplished  in our model by assuming that the two electrons  meet
at a point  modeled by a quantum dot with two spin states
(``up" and ``down").\cite{comAAx}  The Pauli principle is assumed to be effective only there;
it is taken into account by projecting the spin state of one electron on the
spin-up state  of the dot and that
of the other
on the spin-down state there.
The projection
breaks the coherent evolution of the
spin states,
which we refer to as  ``Pauli quenching" of the spin precession.
As the electrons leave the meeting point
their spin states again evolve coherently. In this sense the coherent propagation of the electrons through the link is broken up into
two pair-hopping events.
The Coulomb interaction in the d-channel
is accounted for only at the meeting point. Assuming  that
 the electrons are equally likely to meet anywhere along the
nanowire, the
 outcome of this event for a particular
choice of  meeting
location
is averaged over all possible
choices.

Below, we first introduce
the Hamiltonian of our model and detail the calculation of the transmission
of Cooper pairs between two superconductors connected by an SOI-active weak link. Next,
the spin-precession factor of each of the two processes is presented, and  the way  the disparity between the two reflects the coherence of the s-channel process, and the incoherence of the d-channel one is explained.  Explicit expressions for these   factors,  for a specific model of the Rashba linear SOI,  are then analyzed, followed by a discussion of relevant experiments.

\noindent{\bf The model and the current.}
As mentioned, the ``meeting point" of the two electrons  is represented by a single-level quantum dot of energy $\epsilon$, that  can  accommodate  the transferred electrons in ``up"  and  ``down"  spin states.
The passage of the electrons in and out of the dot is viewed as  single-electron tunneling events, whose amplitudes include the electronic spin precession;
the reservoirs that supply the electrons are two bulk BCS superconductors,
coupled together by a nanowire on which the quantum dot is located.
Thus, the Hamiltonian of the entire junction reads
\begin{align}
{\cal H}={\cal H}^{}_{0}+{\cal H}^{}_{\rm tun}\ ,
\end{align}
where ${\cal H}_{0}$
describes the decoupled system,
the Hamiltonian of the quantum dot and that of the
leads,
\begin{align}
{\cal H}^{}_{0}=
\sum_{\sigma}\epsilon d^{\dagger}_{\sigma}d^{}_{\sigma}+U
d^{\dagger}_{\up}d^{}_{\up}d^{\dagger}_{\down}d^{}_{\down}
+\sum_{\alpha=L,R}{\cal H}^{\alpha}_{\rm lead}\ .
\end{align}
Here,
$d^{}_{\sigma}$ ($d^{\dagger}_{\sigma}$) annihilates (creates) an electron in the spin state $|\sigma\rangle $ on the dot and  $U$ denotes the Coulomb repulsion energy.
The BCS  leads
are described  by the annihilation (creation) operators  of the electrons there, $c^{}_{{\bf k}({\bf p})\sigma}$ ($c^{\dagger}_{{\bf k}({\bf p})\sigma}$).
[${\bf k}$ (${\bf p}$) enumerates the single-particle orbital states  on the left (right) lead.] Denoting by  $
\epsilon^{}_{k(p)}$  the single-electron energy measured relative to the chemical potential,  \cite{com1}
the Hamiltonian of the leads is
\begin{align}
&{\cal H}^{\alpha=L(R)}_{\rm lead}=\sum_{{\bf k} ({\bf p}),\sigma}\epsilon^{}_{ k(p)}c^{\dagger}_{{\bf k}({\bf p})\sigma}c^{}_{{\bf k}({\bf p})\sigma}\nonumber\\
&-\Delta^{}_{L(R)}
\sum
_{{\bf k}({\bf p})}(e^{i\phi^{}_{L(R)}}c^{\dagger}_{{\bf k}({\bf p})\up}c^{\dagger}_{-{\bf k}(-{\bf p})\down}+{\rm H.c.})\ ,
\label{Hl}
\end{align}
where
$\Delta_{L(R)}$ and $\phi_{L(R)}$ are
the amplitude and the phase of the superconducting order parameters.

The tunneling Hamiltonian is the key component of our model,
\begin{align}
{\cal H}^{}_{\rm tun}={\cal H}^{}_{LD}+{\cal H}^{}_{RD}+{\cal H}^{}_{DL}+{\cal H}^{}_{DR}\ .
\label{Htun1}
\end{align}
The transfer of an electron from the spin state $|\sigma'\rangle$ on the  dot to the state $|{\bf k}({\bf p}),\sigma\rangle$ in the left (right) reservoir
is expressed by
\begin{align}
{\cal H}^{}_{L(R)D}&=\sum_{{\bf k}({\bf p}),\sigma,\sigma'}[t^{}_{{\bf k}({\bf p})}]^{}_{\sigma\sigma'}c^{\dagger}_{{\bf k}({\bf p})\sigma}d^{}_{\sigma'}
\ ,
\label{Htun2}
\end{align}
while the reverse  process,   from the  state $|{\bf k}({\bf p}),\sigma\rangle $  in the left (right) lead to the spin state $|\sigma'\rangle $ on the dot, is ${\cal H}^{}_{DL(R)}=[{\cal H}^{}_{L(R)D}]^\dagger$.
The amplitude
$[t^{}_{{\bf k}({\bf p}) }]_{\sigma\sigma'}$ allows for  spin flips during the tunneling.
It is conveniently separated
into a (scalar) orbital amplitude, and a matrix that contains the
effects of the SOI (whether
of the Rashba \cite{Rashba.1960,Bychkov.1984} or the Dresselhaus \cite{Dresselhaus.1955} type), and also 
the dependence on the
spatial direction of the SOI-active wire.
For
the linear  SOI  \cite{Shekhter.2017}
\begin{align}
{\bf t}^{LD}_{\bf k}=it^{}_{L}e^{-ik^{}_{\rm F}d^{}_{L}}{\bf W}^{LD}_{}\ ,
\label{GF}
\end{align}
where $k_{\rm F}$ is the Fermi wave vector in the leads,  and  $d_{L}$
is the length of the bond between
the left lead and the dot. The superscript  indicates the tunneling direction, from the dot to the left  lead, {\it etc}.
Specific forms for  ${\bf W}$ are discussed below.

Since the two electrons of a Cooper pair are in time-reversed states we also have to consider the transfer of an electron from
the time-reversed spin state $\vert\overline{\sigma}'\rangle \equiv (i\sigma_y)\vert\sigma'\rangle$ on the dot to the state
$\vert -{\bf k}(-{\bf p}),\overline{\sigma}\rangle$ in the left (right) lead.
The amplitude for this process is $[\overline{t}^{
}_{{\bf k}({\bf p})}]_{\overline{\sigma}\overline{\sigma}'}$,
where
\begin{align}
\label{tbar}
\overline{\bf t}^{L(R)D
}_{{\bf k}({\bf p})} \equiv  \hat{\bf  T} {\bf t}^{L(R)D
}_{{\bf k}({\bf p})} \hat{\bf T}^{-1}_{}
\,; \quad \hat{\bf  T}=K (i\sigma^{}_y)\ ,
\end{align}
($\sigma_y$ is a Pauli matrix,  and $K$ is the complex conjugation operator). We
note here the important relation
\begin{align}
[\overline{t}^{DL(R)
}_{{\bf k}({\bf p})}]_{\overline{\sigma}\overline{\sigma}'}=[t^{DL(R)
\ast}_{{\bf k}({\bf p})}]_{\sigma\sigma'}\ ,
\label{TR}
\end{align}
used below
to eliminate the time-reversed tunneling amplitudes from our final results.

The flow of electrons between the two superconductors
is analyzed by studying the equilibrium Josephson current, i.e.,  the rate by which electrons leave the left superconductor  \cite{com2}  (we use $\hbar=1$)
\begin{align}
J^{}_{L}=-e (d/dt)\langle\sum_{{\bf  k},\sigma}
c^{\dagger}_{{\bf k}\sigma}c^{}_{{\bf k}\sigma}\rangle=-2e\,{\rm Im}\langle {\cal H}^{}_{LD}(t)\rangle\ ,
\end{align}
where the angular brackets denote quantum averaging.
$J_{L}$ is evaluated
using the $S-$matrix, $\langle {\cal H}^{}_{LD}(t)\rangle = \langle S^{-1}(t,-\infty) {\cal H}^{}_{LD}(t) S(t, -\infty) \rangle$,
with ${\cal H}_{LD}(t) = \exp[i{\cal H}_{0}t] {\cal H}_{LD} \exp [-i{\cal H}_{0}t]$, and  the quantum average is
 with respect to ${\cal H}_0$.
As it  is at least fourth-order in the tunneling Hamiltonian,
it is found from the expansion up to  third order of the $S$-matrix.    \cite{Gisselfalt}
The energy level on the dot is assumed to lie well above the chemical potential 
of the leads,  and thus the small parameter of the expansion is $\Gamma/\epsilon$, where
$\Gamma=\Gamma_{L}^{}+\Gamma^{}_{R}$
is the width of the resonance level created on the dot due to the coupling with the bulk reservoirs. This implies that the perturbation expansion is carried out on a dot which is initially empty. \cite{occupied.dot}

The lowest-order
current
results from the processes in  which two electrons are injected into  and extracted from the dot.
Two groups of terms can be identified. In
the first double occupancy on the dot does not occur,  and the transfer of the electron pair  is accomplished by a sequential tunneling of the paired electrons one by one.
These terms, which form the  s-channel,  contain
$\langle d^{}_{\sigma_{i}}(t^{}_{i})d^{\dagger}_{\sigma^{}_{j}}(t^{}_{j})d^{}_{\sigma'^{}_{i}}(t^{}_{i'})d^{\dagger}_{\sigma'_{j}}(t^{}_{j'})\rangle$.
The Pauli exclusion principle is not active here;
the  spin states are determined such that  the initial and final states of the dot are empty
(i.e., $\sigma'_{i}=\sigma'_{j}$ and $\sigma_{i}=\sigma_{j}$). The contribution to the Josephson current from these
is \cite{com3}
\begin{align}
J^{\rm s}_{}=I^{}_{0}
F^{\rm s}_{}(\epsilon/\Delta ){\cal A}^{\rm s
}_{}\ ,
\label{Jsin}
\end{align}
where  $\Delta_{L}=\Delta_{R}=\Delta$  is assumed.  For a short  link the prefactor is \cite{Glazman-Matveev}
$I^{}_{0}=
2e[\Gamma^{}_{L}\Gamma^{}_{R}/\Delta]\sin (\phi^{}_{R}-\phi^{}_{L})$,
and
\begin{align}
F^{\rm s}_{}(\widetilde{\epsilon})
&=\int_{-\infty}^{\infty}(d\zeta^{}_{k}/\pi)
\int_{-\infty}^{\infty}(d\zeta^{}_{p}/\pi)
[({\rm cosh}\zeta^{}_{k}+\widetilde{\epsilon})\nonumber\\
&\times({\rm cosh}\zeta^{}_{k}+{\rm cosh}\zeta^{}_{p})({\rm cosh}\zeta^{}_{p}+\widetilde{\epsilon})]^{-1}\ .
\label{Fsin}
\end{align}
The spin-precession factor  ${\cal A}^{\rm s}$  in
Eq.~(\ref{Jsin})
is given
in  Eq.~(\ref{As1}) and discussed below. 
In the second group of terms  the dot is doubly occupied during the tunneling; these terms, of the generic form
$\langle d^{}_{\sigma_{i}}(t^{}_{i})d^{}_{\sigma^{}_{j}}(t^{}_{j})d^{\dagger}_{\sigma'^{}_{i}}(t^{}_{i'})d^{\dagger}_{\sigma'_{j}}(t^{}_{j'})\rangle$,
constitute the d-channel.
The Pauli exclusion-principle
constrains the spin states on the dot,
$\sigma'_{i}=-\sigma'_{j}$ and $\sigma_{i}=-\sigma_{j}$.
They contribute
\begin{align}
J^{\rm d}_{}=I^{}_{0}
F^{\rm d}_{}(\epsilon/\Delta,U/\Delta){\cal A}^{\rm d}_{}\ ,
\label{Jdou}
\end{align}
where  ${\cal A}^{d}$ is the spin precession factor of these
processes [Eq.~(\ref{Ad1})] and
\begin{align}
F^{\rm d}_{}(\widetilde{
\epsilon},\widetilde{U})
&=\int_{-\infty}^{\infty}(d\zeta^{}_{k}/\pi)
\int_{-\infty}^{\infty}d\zeta^{}_{p}/\pi)
[({\rm cosh}\zeta^{}_{k}+\widetilde{\epsilon})\nonumber\\
&\times(2\widetilde{\epsilon}+\widetilde{U})({\rm cosh}\zeta^{}_{p}+\widetilde{\epsilon})]^{-1}\ .
\label{Fdoub}
\end{align}
Both ${\cal A}^{\rm s}=1$ and ${\cal A}^{\rm d}=1$ in the absence of the SOI.

The complete Josephson current
is the sum of the contributions from the two types of processes, $J=J^{s}+J^{d}$,
\begin{align}
\label{Jtot}
&J=
J^{}_{0}\frac{{\cal A}^{\rm s}_{}
F^{\rm s}_{} (\epsilon/\Delta )+2{\cal A}^{\rm d}_{}F^{\rm d}_{} (\epsilon/\Delta, U/\Delta )}{F^{\rm s}_{} (\epsilon/\Delta )+2F^{\rm d}_{} (\epsilon/\Delta, U/\Delta )}\ ,
\end{align}
where
$J_{0}=I^{}_{0} [
F^{\rm s}_{} (\epsilon/\Delta )+2F^{\rm d}_{} (\epsilon/\Delta, U/\Delta)]$
is  the equilibrium Josephson current of a junction with no SOI. Note that $J$  depends on the location of the dot, via the spin-precession factors ${\cal A}^{\rm s}$ and ${\cal A}^{\rm d}$. \cite{comRS} The Coulomb repulsion on the dot, $U $, affects the relative weights  of the contributions from  the s-  and  the d-channels:
although  $F^{\rm s}$ and $F^{\rm d}$   are  independent of the SOI strength, they  weigh differently the spin-precession  factors.
This competition modifies significantly the current.

\noindent{\bf The spin precession.}
In terms of the SOI amplitudes, Eq.~(\ref{GF}), and using the symmetry relation Eq.~(\ref{TR}),  the spin-precession factor ${\cal A}^{\rm s}$ of the s-channel
 is \cite{com4}
\begin{align}
{\cal A}^{\rm s}_{}&
=\vert W^{LR}_{\up\up}|^{2}_{} - \vert W^{LR}_{\up\down}|^{2}_{}
\ .
\label{As1}
\end{align}
$W^{LR}_{\sigma_{L}\sigma_{R}}\equiv
\sum_{\sigma}W^{LD}_{\sigma^{}_{L}\sigma}\,W^{DR}_{\sigma\sigma^{}_{R}}$ is
the
direct tunneling amplitude
between the leads. \cite{com5}  As opposed, the spin precession factor of the d-channel cannot be expressed in terms of the direct  amplitudes.
Using Eqs.~(\ref{GF}) and (\ref{TR}) we find
\begin{align}
\label{Ad1}
&{\cal A}^{\rm d}_{}
=\left(\vert W^{LD}_{\up\up}|^{2}_{} - \vert W^{LD}_{\up\down}\vert^2\right) \left(\vert W^{DR}_{\up\up}|^{2}_{} - \vert W^{DR}_{\up\down}\vert^2\right) \ .
\end{align}
It is interesting to compare the structure of Eqs.~(\ref{As1}) and (\ref{Ad1}). \cite{comAA2} One notes that ${\cal A}^{\rm d}$ is a product of two factors of the same structure as the single factor in ${\cal A}^{\rm s}$.
Our interpretation is that ${\cal A}^{\rm s}$ describes the coherent transfer of a Cooper pair from the right to the left lead, while
${\cal A}^{\rm d}$ describes  first  a coherent Cooper pair transfer from the right lead to the dot, where coherence is lost, then a second
coherent transfer from the dot to the left lead.

\noindent{\bf Linear spin-orbit couplings.}
Though it is possible to calculate an effective SOI  {\it ab initio},  it is convenient to adopt the phenomenological Rashba Hamiltonian,
\cite{Bychkov.1984}
valid for systems with a single high-symmetry axis that lack spatial inversion symmetry. For  an electron of an effective mass $m^{\ast}$ and  momentum ${\bf p}$ propagating along  a wire where the SOI is active, it reads
${\cal H}_{\rm so}=(\hbar k^{}_{\rm so}/m^{\ast})\sig\cdot({\bf p}\times\hat{\bf n})$,
where   $\hat{\bf n}$ is a unit vector along the symmetry axis (the $\hat{\bf c}-$axis in  hexagonal wurtzite crystals, the growth direction in a semiconductor heterostructure, the direction of an external electric field),  and $k_{\rm so}$ is the strength of the SOI  in units of inverse length. Using  this Hamiltonian,  we find  \cite{Shekhter.2013}
\begin{align}
{\bf t}^{}_{{\bf k}({\bf p})}= it^{}_{L}e^{ik^{}_{\rm F}d^{}_{L(R)}}\exp[ik^{}_{\rm so}{\bf d}^{}_{L(R)}\times\hat{\bf n}\cdot\sig]\ ,
\label{GF1}
\end{align}
where ${\bf d}_{L(R)} $ is the radius vector pointing from the dot to the left (right) reservoir  along the wire. The linear Dresselhaus SOI \cite{Dresselhaus.1955} leads to a similar form.  \cite{Ora.2005,Rashba.2010}

\begin{figure}
\includegraphics[width=8cm]{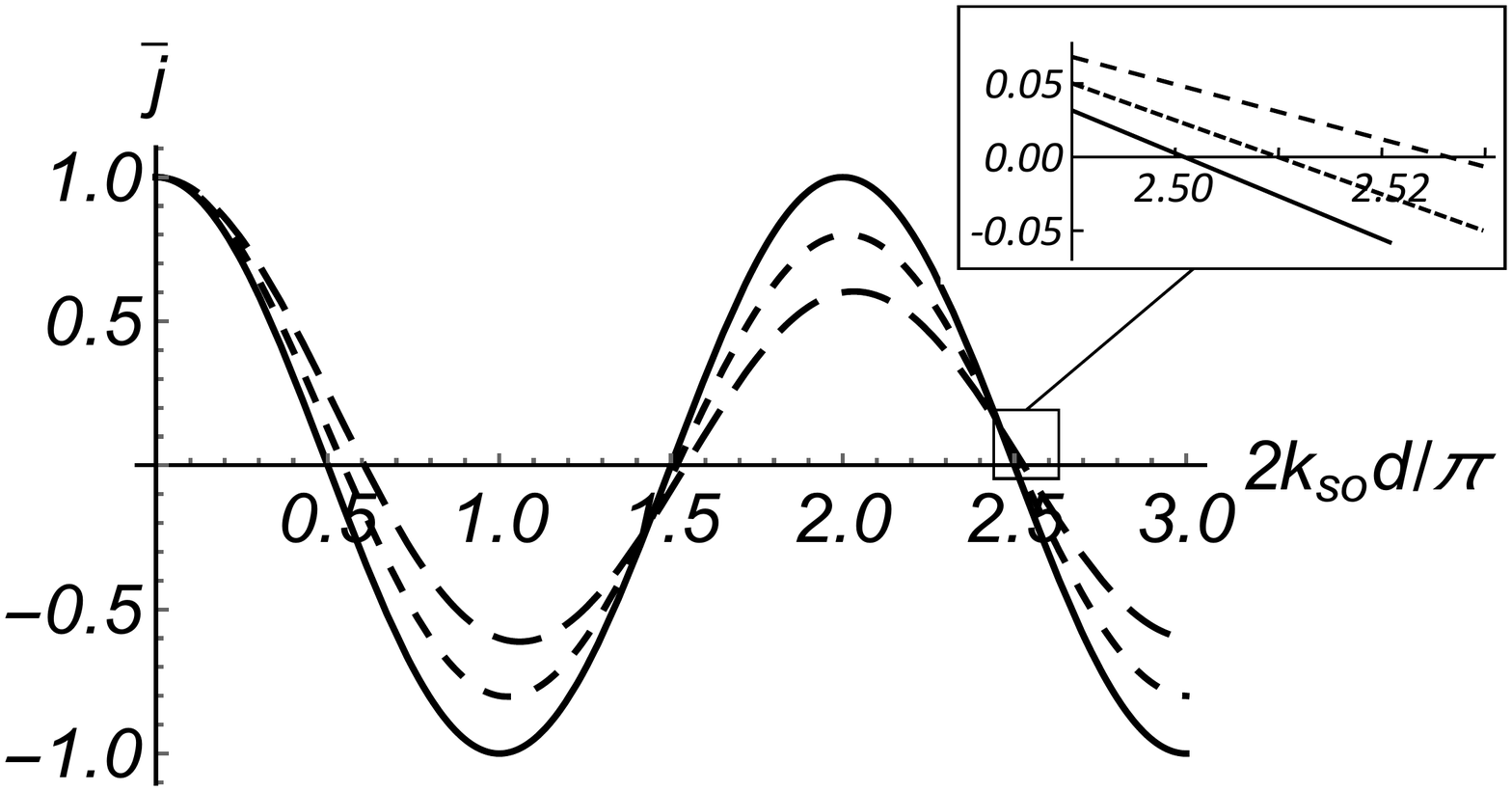}
\vspace{-1.3cm}
\caption{The average Josephson current $\bar{j}$ {\it vs.} $2k_{\rm so}d/\pi$, for various values of the Coulomb repulsion, as encoded in $Z$: $Z=0$ (full line) and  $Z=0.2,\ 0.4$ (increasing dashes).  Inset:
a narrower range, where $\bar{j}$ changes sign.
}
\label{Fig}
\end{figure}

As an explicit example, we consider a straight  nanowire \cite{com6} of length $d$ as a weak link, on which the electrons are subjected to the Rashba SOI.  The wire lies  along  $\hat{\bf x}$ in  the $XY$ plane, and is perpendicular to $\hat{\bf n}$.  Then
\begin{align}
{\cal A}^{\rm s}_{}&= \cos(2k^{}_{\rm so}d)\ ,\ {\cal A}^{\rm d}_{}=
\cos(2k^{}_{\rm so}d^{}_{L})
\cos(2k^{}_{\rm so}d^{}_{R})\ ,
\label{AsAd}
\end{align}
which reflects the structure of  Eqs.~(\ref{As1}) and (\ref{Ad1}).
The spin-precession  of the s-channel
is independent of the position of the dot.
Placing the dot at a distance  $x$ from the left reservoir, the spin-precession  of the d-channel is ${\cal A}^{\rm d}=\cos(2k^{}_{\rm so}d)+\sin[2k^{}_{\rm so}(d-x)]\sin(2k^{}_{\rm so}x)$.  Averaging the normalized current over $x$ yields  \cite{comAAx}
\begin{align}
&\bar{j}=
\int_{0}^{d}\frac{dx}{d}
\frac{J(x)}{J^{}_{0}}=\Big  [(1-Z)\cos(2k^{}_{\rm so}d)
+Z\frac{\sin(2k^{}_{\rm so}d)}{2k^{}_{\rm so}d}
 \Big ]\ ,\nonumber\\
& Z(U)\equiv{\cal F}^{\rm d}(\frac{\epsilon}{\Delta},\frac{U}{\Delta})\Big/\Big [{\cal F}^{\rm s}(\frac{\epsilon}{\Delta})
+2{\cal F}^{\rm d}(\frac{\epsilon}{\Delta},\frac{U}{\Delta})\Big ] \ .
\label{Jtotav}
\end{align}
In spite of the averaging over the possible locations where the two electrons' wave packets overlap considerably,  the spin-orbit dynamics is preserved: the  current oscillates as a function of the gate voltage which dictates  $k_{\rm so}$.
Figure ~\ref{Fig} shows $\bar{j}$ {\it vs.} $2k_{\rm so}d$ for a range of values
taken from the experiments,  for several values of $Z$.
In the Coulomb-blockade limit  $U=\infty$ and  $Z=0$:  the current exhibits the simple oscillations $\bar{j}=\cos(2k_{\rm so}d)$.   As $U$ decreases, $Z$ increases monotonically (for fixed $\epsilon/\Delta$) but remains smaller than $1/2$. It
also decreases with increasing $\epsilon/\Delta$. At large $2k_{\rm so}d$ $\bar{j}\approx (1-Z)\cos(2k_{\rm so}d)$; the Coulomb interaction  reduces the magnitude of $\bar{j}$, but does not change its sign. For intermediate values of $2k_{\rm so}d$
there appear small segments of $2k_{\rm so}d$ in which the sign of the current also changes for $Z>0$, i.e., for
$\tan(2k_{\rm so}d)/(2k_{\rm so}d)<(Z-1)/Z$.
\ The inset in Fig.~\ref{Fig} zooms in on such a range;  it broadens  as $Z$ increases.

The capability to tune the SOI
electrostatically
by gate voltages
was demonstrated
on the inversion layer of a
In$^{}_{0.75}$Ga$^{}_{0.25}$As/In$^{}_{0.75}$Al$^{}_{0.25}$As semiconductor heterostructure. \cite{Sato.2001}
The spin-orbit coupling constant $\alpha$, mainly attributed to the
Rashba SOI parameter $\alpha_R$, varied with gate voltage between roughly 150 and 300 meV\AA.
Using $k_{\rm so}=m^\ast\alpha_R/\hbar^2$
and the measured
$m^\ast=0.041m$ ($m$ is the mass of a free electron)  one concludes
that if a weak link were to be electrostatically defined in this system the argument $2k_{\rm so}d$ of the trigonometric functions in
 Eq.~(\ref{Jtotav}) for a 1 $\mu$m long link could be varied from $\sim 16$ to $\sim 32$. This amounts to a tuning over a range that is more than twice
the period $2\pi$
of these functions.
A more recent
experimental evidence for the SOI tunability
is found in
a dual gated InAs/GaSb quantum well where the Rashba SOI parameter $\alpha_R$ could be varied
between 53 and 75 meV\AA, while the Dresselhaus SOI was kept constant. \cite{Beukman.2017}   The stated value of $m^\ast=0.04m$ implies that  $2k_{\rm so}d$ could be varied between $\sim 6$ to $\sim 8$, that is over about a third of
the period $2\pi$
if $d=1\, \mu$m.

The magnitude of the Josephson current through a quantum dot is set by the functions ${\cal F}^{\rm s}$ [Eq.~(\ref{Fsin})] and ${\cal F}^{\rm d}$
[Eq.~(\ref{Fdoub})], that
are derived  for  short  weak links. \cite{Glazman-Matveev}  However, whereas
the restriction on the length $d$ of the link might be strict,  $d\ll\xi$,  for  the orbital part ($\xi$ is the superconducting coherence length), it is far weaker for the spin-dependent part:
$k^{}_{\rm so}d\ll k^{}_{\rm F}\xi$, since the  spin-precession factors ${\cal A}^{\rm s}$ and ${\cal A}^{\rm d}$
are not sensitive to the energy dependence of the transmission amplitude. \cite{Shekhter.2017}

\noindent{\bf Summary.}
We have considered
 the spin splitting of Cooper pairs that carry a supercurrent
through a weak-link Josephson junction. Our main result, expressed in Eq.~(\ref{Jtotav}) and shown in Fig.~\ref{Fig}, is that
Coulomb repulsion and   Pauli quenching do
affect the current, but
do not destroy the possibility to tune  it  by changing
the strength of the spin-orbit interaction.

The oscillatory dependence of the supercurrent on the SOI strength
results from a rather complex interference between different  transmission events: the single-electron transmission one (s-channel),  that yields
$J^{\rm s}$, Eq.~(\ref{Jsin}), and the double-electron transmission
(d-channel) that gives
$J^{\rm d}$, Eq.~(\ref{Jdou}). In the s-channel the two electrons are transferred one by one, so
that at any time during the tunneling there is only one electron in the  link. By contrast,
in the d-channel both electrons appear in the  link for some period of time, which means that in
the Coulomb blockade limit the transfer of Cooper pairs in this channel is completely suppressed. 
The s-channel  has two coherent transmission
channels, one where the spins of both members of the pair are preserved [first term in Eq. (\ref{As1})]
and one where they are both flipped (second term there).  This double spin reversal is
equivalent to a permutation of the paired electrons, which explains
the difference in sign between the two terms.
As the Coulomb blockade is lifted,  the probability of pairs to be transferred in
the d-channel increases.  As  seen from Eq. (\ref{Ad1}) for the spin precession factor ${\cal A}^{\rm d}$,
the d-channel  involves coherent transfers of the pairs separated by a Pauli quenching that
breaks coherence. Remarkably enough,  each of the two factors of  ${\cal A}^{\rm d}$, which describes two coherent ``hops", have the same structure as the result ${\cal A}^{\rm s}$.

The pronounced oscillations of the supercurrent
and the sign reversal can be observed for plausible  lengths of the weak link, of the order of a  micron,  supposedly achievable by suitably-designed geometries of the gates.
This result indicates
interesting phenomena caused by  SOI-induced spin polarization of Cooper pairs.

The model, suggested here, allows for a unified approach to treat an interplay between spin- and charge - related phenomena, which are in the center of nowadays nanoelectronics. An immediate generalization of this approach would be important for exploring new device functionalities such
as spin-orbit controlled shot noise in nanostructures and spin-orbit effect in current cross-correlations in multi-terminal nanodevices.


We thank the Computational Science Research Center in Beijing  for the hospitality that allowed for the accomplishment of this project. RIS and MJ thank the IBS Center for Theoretical Physics of Complex Systems, Daejeon, Rep. of Korea, and OEW and AA thank the Dept. of Physics, Univ. of Gothenburg, for hospitality. This work was partially supported by the Swedish Research
Council (VR), by the Israel Science Foundation
(ISF) and by the infrastructure program of Israel
Ministry of Science and Technology under contract
3-11173.



\begin{thebibliography}{99}
\bibitem{comAAx}
Since we cannot know exactly where the two electrons
``meet", we average over all the possible location of this ``quantum dot".
\bibitem{com1}
As we focus on equilibrium transfers of pairs of electrons, it is assumed tacitly  that the chemical potentials of the leads are identical. Under these circumstances, our calculation produces just the equilibrium Josephson current.
\bibitem{occupied.dot}
The other possibility,  of a dot which is initially occupied, is also of interest. We hope to pursue the spin-splitting phenomenon for this configuration in the future.
\bibitem{com3}
For simplicity, the expression for the Josephson current is presented at zero temperature; the extension to finite temperatures is quite straightforward. The subscript $L$ on the current  is not needed, as at equilibrium $J_{L} =  J_{R}$.
\bibitem{comRS}
We postulate in our model that current $J_{0}$  defined by Eqs. (\ref{Jsin}) and (\ref{Jdou}) does not depend on the location  of the dot.
\bibitem{com6}
Equations (\ref{As1}) and (\ref{Ad1}) for
${\cal A}^s$ and ${\cal A}^d$ are valid for straight and bent weak links. For simplicity, we presented numerical results only for  the simpler case of a straight link, for which ${\cal A}^s$ and ${\cal A}^d$ are
given by Eq. (\ref{AsAd}).
\end{thebibliography}
\end{document}